\documentstyle[11pt]{article}
\begin{document}
\begin{titlepage}
\begin{center}
{\Large \bf ``Finite" Non-Gaussianities and Tensor-Scalar Ratio in Large Volume Swiss-Cheese Compactifications}
\vskip 0.1in { Aalok Misra\footnote{e-mail: aalokfph@iitr.ernet.in} and
Pramod Shukla \footnote{email: pmathdph@iitr.ernet.in}\\
Department of Physics, Indian Institute of Technology,
Roorkee - 247 667, Uttaranchal, India}
\vskip 0.5 true in
\end{center}
\thispagestyle{empty}
\begin{abstract}
Developing on the ideas of (section 4 of) \cite{dSetal} and \cite{axionicswisscheese} and using the formalisms of \cite{Yokoyamafnl,Yokoyama}, after inclusion of perturbative and non-perturbative $\alpha^\prime$ corrections to the K\"{a}hler potential and ($D1$- and $D3$-) instanton generated superpotential, we show the possibility of getting finite values for the non-linear parameter $f_{NL}$ while looking for non-Gaussianities in type IIB compactifications on orientifolds of the Swiss Cheese Calabi-Yau ${\bf WCP}^4[1,1,1,6,9]$ in the L(arge) V(olume) S(cenarios) limit. First we show that in the context of multi-field slow-roll inflation, for the Calabi-Yau volume ${\cal V}\sim{10}^5$ and $D3-$instanton number $n^s\sim10$ along with $N_e\sim18$, one can realize $f_{NL}\sim0.03$, and for Calabi-Yau volume ${\cal V}\sim{10}^6$ with $D3-$instanton number $n^s\sim10$ resulting in number of e-foldings $N_e\sim{60}$, one can realize $f_{NL}\sim0.01$. Further we show that with the slow-roll conditions violated and for the number of the $D3-$instanton wrappings $n^s\sim{\cal O}(1)$, one can realize $f_{NL}\sim{\cal O}(1)$. Using general considerations and some algebraic geometric assumptions, we show that with requiring a ``freezeout" of curvature perturbations at super horizon scales, it is possible to get tensor-scalar ratio $r\sim{\cal O}(10^{-3})$ with  the loss of scale invariance $|n_R-1|=0.01$ and one can obtain $f_{NL}\sim{\cal O}({10}^{-2})$ as well in the context of slow-roll inflation scenarios in the same Swiss-Cheese setup. For all our calculations of the world-sheet instanton contributions to the K\"{a}hler potential coming from the non-perturbative $\alpha^\prime$ corrections,  the degrees of genus-zero rational curves  correspond to the largest value of the Gopakumar-Vafa invariants for the chosen compact projective variety, which is very large. To our knowledge, such values of non-Gaussianities and tensor-scalar ratio in slow-roll inflationary and/or slow-roll violating scenarios, have been obtained for the first time from string theory. We also make some observations pertaining to the possibility of the axionic inflaton also being a cold dark matter candidate as well as a quintessence field used for explaining dark energy.
\end{abstract}
\end{titlepage}

\section{Introduction}

The idea of inflation has widely involved several string theorists for the past few years with attempts to constructing realistic inflationary models to have a match with the cosmological observations \cite{Bartolotheoryobservations,Bartolocmbaniso,Hajiancmb_data,Kogo,CMb} as a test of string theory \cite{kallosh,KKLMMT,kallosh1,AxionInflation,wmap}. Although the idea of inflation was initially introduced to explain the homogeneous and isotropic nature of the universe at large scale structure \cite{FirstInflation,cosmoproblem,linde}, its best advantage is reflected while studying inhomogeneities and anisotropries of the universe, which is a consequence of the vacuum fluctuations of the inflaton as well as the metric fluctuations. These fluctuations result in non-linear effects (parametrized by $f_{NL},\tau_{NL}$) seeding the non-Gaussianity of the primordial curvature perturbation, which are expected to be observed by PLANCK, with non-linear parameter $f_{NL} \sim {\cal O}(1)$ \cite{Planck}. Along with the non-linear parameter $f_{NL}$, the ``tensor-to-scalar ratio" $r$ is also one of the key inflationary observables, which measures the anisotropy arising from the gravity-wave(tensor) perturbations and the signature of the same is expected to be predicted by the PLANCK if the tensor-to-scalar ratio $r \sim 10^{-2}-10^{-1}$ \cite{Planck}. As these parameters give a lot of information about the dynamics inside the universe, the theoretical prediction of large/finite (detectable) values of the non-linear parameters $f_{NL},\tau_{NL}$ as well as ``tensor-to-scalar ratio" $r$ has received a lot of attention for recent few years \cite{Hajiancmb_data,Kogo,Alabidi,Byrnes_Wands,Byrnestrispectrum,alishahiha,beyondslowroll,kinney2,finite_r1,
finite_r,slowrollflow,slowflowturner,kinney3,Maldacenadeltan}. For calculating the non-linear parameter $f_{NL}$, a very general formalism (called as ${\delta N}$ -formalism) was developed and applied for some models \cite{deltaN}. Initially the parameter $f_{NL}$ was found to be suppressed (to undetectable value) by the slow roll parameters in case of the single inflaton model. Followed by this, several models with multi-scalar fields have been proposed but again with the result of the non-linear parameter $f_{NL}$ of the order of the slow-roll parameters as long as the slow-roll conditions are satisfied \cite{Yokoyamafnl,Yokoyama,Bartolotheoryobservations,Sasaki,Rigopoulosnlpertb,Rigopouloslargenongaussianity,Rigopoulosbispectra,Seerylidsey,Battefeldmultifield,Choihall,Wandstwofield}. Recently considering multi-scalar inflaton models, Yokoyama et al have given a general expression for calculating the non-linear parameter $f_{NL}$ (using ${\delta N}$ -formalism) for non-separable potentials\cite{Yokoyamafnl} and found the same to be suppressed again by the slow-roll parameter $\epsilon$ (with an enhancement by exponential of quantities ${\sim {\cal O}(1)}$). In the work followed by the same as a generalization to the non slow-roll cases, the authors have proposed a model for getting finite $f_{NL}$  violating the slow roll conditions temporarily \cite{ Yokoyama}. The observable ``tensor-to-scalar ratio" $r$, characterizing the amount of anisotropy arising from scalar-density perturbations (reflected as the CMB quadrupole anisotropy) as well as the gravity-wave perturbations arising through the tensorial metric fluctuations, is crucial for the study of temperature/angular anisotropy from the CMB observations. The ``tensor-to-scalar ratio" $r$ is defined as the ratio of squares of the amplitudes of the tensor to the scalar perturbations defined through their corresponding power spectra. Several efforts have been made for getting large/finite value of `` $r$ " with different models, some resulting in small undetectable values while some predicting finite bounds for the same \cite{Alabidi,kinney2,finite_r1,finite_r,slowrollflow,slowflowturner,kinney3}.

In this note, continuing with the results of our previous paper \cite{axionicswisscheese} for Large Volume multi-axionic Swiss-Cheese inflationary scenarios, we discuss whether it is possible, {\it starting from string theory}, to
\begin{itemize}
\item
obtain $f_{NL}\sim{\cal O}({10}^{-2}-{10}^{0})$ ,

\item
obtain tensor-scalar ratio $r\sim10^{-3}$,

\item
obtain number of e-foldings $N_e\sim{\cal O}(10)$,

\item
obtain the loss of scale invariance within experimental bound: $|n_R-1|\le{0.05}$

such that

\item
the curvature perturbations are ``frozen" at super horizon scales,

\item
the inflaton could be a dark matter candidate at least in some corner of the moduli space,

\item
the inflaton could also be identified with quintessence to explain dark energy, once again, at least in some corner - the same as above - of the moduli space.

\end{itemize}
We find that the answer is a yes. The crucial input from algebraic geometry that we need is the fact that Gopakumar-Vafa invariants of genus-zero rational curves for compact Calabi-Yau three-folds expressed as projective varieties can be very large for appropriate maximum values of the degrees of the rational curves. This is utilized when incorporating the non-perturbative $\alpha^\prime$ contribution to the K\"{a}hler potential.

The plan of the paper is as follows. In section {\bf 2}, we review our previous work pertaining to obtaining meta-stable dS vacua without addition of anti-D3 branes and obtaining axionic slow-roll inflation. In section {\bf 3}, using the techniques of \cite{Yokoyamafnl,Yokoyama}, we show the possibility of getting finite ${\cal O}({10}^{-2})$ non-Gaussianities in slow-roll and ${\cal O}(1)$ non-Gaussianities in slow-roll violating scenarios. In section {\bf 4}, based on general arguments not specific to our (string-theory) set-up and using the techniques of \cite{beyondslowroll,large_eta_Kinney}, we show that ensuring ``freezeout" of curvature perturbations at super horizon scales, one can get a tensor-scalar ratio  $r\sim{\cal O}(10^{-3})$ in the context of slow-roll scenarios. In section {\bf 5}, we summarize our results and give some arguments to show the possibility of identifying the inflaton, responsible for slow-roll inflation, to also be a dark matter candidate as well as a quintessence field for sub-Planckian axions.

\section{Review of Large Volume Slow-Roll Axionic Inflationary Scenarios Including Non-Perturbation $\alpha^\prime$ Corrections}

Let us first summarize the results of our previous works (section {\bf 4} of \cite{dSetal}) and \cite{axionicswisscheese}. With the inclusion of perturbative (using \cite{BBHL}) and non-perturbative (using \cite{Grimm}) $\alpha^\prime$-corrections as well as the loop corrections (using \cite{loops}), the K\"{a}hler potential for the two-parameter ``Swiss-Cheese" Calabi-Yau expressed as a projective variety in ${\bf WCP}^4[1,1,1,6,9]$, can be shown to be given by:
\begin{eqnarray}
\label{eq:nonpert81}
& & K = - ln\left(-i(\tau-{\bar\tau})\right) -ln\left(-i\int_{CY_3}\Omega\wedge{\bar\Omega}\right)\nonumber\\
 & & - 2\ ln\Biggl[{\cal V} + \frac{\chi(CY_3)}{2}\sum_{m,n\in{\bf Z}^2/(0,0)}
\frac{({\bar\tau}-\tau)^{\frac{3}{2}}}{(2i)^{\frac{3}{2}}|m+n\tau|^3}\nonumber\\
& & - 4\sum_{\beta\in H_2^-(CY_3,{\bf Z})} n^0_\beta\sum_{m,n\in{\bf Z}^2/(0,0)}
\frac{({\bar\tau}-\tau)^{\frac{3}{2}}}{(2i)^{\frac{3}{2}}|m+n\tau|^3}cos\left((n+m\tau)k_a\frac{(G^a-{\bar G}^a)}{\tau - {\bar\tau}}
 - mk_aG^a\right)\Biggr]\nonumber\\
 & & +\frac{C^{KK\ (1)}_s(U_\alpha,{\bar U}_{\bar\alpha})\sqrt{\tau_s}}{{\cal V}\left(\sum_{(m,n)\in{\bf Z}^2/(0,0)}\frac{\frac{(\tau-{\bar\tau})}{2i}}{|m+n\tau|^2}\right)} + \frac{C^{KK\ (1)}_b(U_\alpha,{\bar U}_{\bar\alpha})\sqrt{\tau_b}}{{\cal V}\left(\sum_{(m,n)\in{\bf Z}^2/(0,0)}\frac{\frac{(\tau-{\bar\tau})}{2i}}{|m+n\tau|^2}\right)}.
\end{eqnarray}
In (\ref{eq:nonpert81}), the first line and $-2\ ln({\cal V})$ are the tree-level contributions. The second (excluding the volume factor in the argument of the logarithm) and third lines are the perturbative and non-perturbative $\alpha^\prime$ corrections. $\{n^0_\beta\}$ are the genus-zero Gopakumar-Vafa invariants that count the number of genus-zero rational curves. The fourth line is the $1$-loop contribution; $\tau_s$ is the volume of the ``small" divisor and $\tau_b$ is the volume of the ``big" divisor. The loop-contributions arise from KK modes corresponding to closed string or 1-loop open-string exchange between $D3$- and $D7$-(or $O7$-planes)branes wrapped around the ``s" and ``b" divisors. Note that the two divisors for
${\bf WCP}^4[1,1,1,6,9]$, do not intersect (See \cite{Curio+Spillner}) implying that there is no contribution from winding modes corresponding to strings winding non-contractible 1-cycles in the intersection locus corresponding to stacks of intersecting $D7$-branes wrapped around the ``s" and ``b" divisors. One sees from (\ref{eq:nonpert81}) that in the LVS limit, loop corrections are sub-dominant as compared to the perturbative and non-perturbative $\alpha^\prime$ corrections.

To summarize the result of section 4 of \cite{dSetal}, one gets the following potential:
\begin{eqnarray}
\label{eq:nonpert21}
& & V\sim\frac{{\cal Y}\sqrt{ln {\cal Y}}}{{\cal V}^{2n^s+2}}e^{-2\phi}\frac{\left(\sum_{n^s}n^s\sum_{m^a}{e^{-\frac{m^2}{2g_s} + \frac{m_ab^a n^s}{g_s} + \frac{n^s\kappa_{1ab}b^ab^b}{2g_s}}}\right)^2}{\left|f(\tau)\right|^2}
\nonumber\\
& & + \sum_{n^s}\frac{W ln { \cal Y} }{{\cal V}^{n^s+2}}\left(\frac{\theta_{n^s}({\bar\tau},{\bar G})}{f(\eta({\bar\tau}))}
\right)e^{-in^s(-\tilde{\rho_1}+\frac{1}{2}\kappa_{1ab}
\frac{{\bar\tau}G^a-\tau{\bar G}^a}{({\bar\tau}-\tau)}\frac{(G^b-{\bar G}^b)}{({\bar\tau}-\tau)} -
\frac{1}{2}\kappa_{1ab}\frac{G^a(G^b-{\bar G}^b)}{(\tau-{\bar\tau})})}+c.c.\nonumber\\
& & +
\sum_{k^1,k^2}\frac{|W|^2}{{\cal V}^3}\left(\frac{3k_2^2+k_1^2}{k_1^2-k_2^2}\right)
\frac{\left|\sum_c\sum_{n,m\in{\bf Z}^2/(0,0)}e^{-\frac{3\phi}{2}}A_{n,m,n_{k^c}}(\tau) sin(nk.b+mk.c)\right|^2}
{\sum_{c^\prime}\sum_{m^\prime,n^\prime\in{\bf Z}^2/(0,0)} e^{-\frac{3\phi}{2}}|n+m\tau|^3
|A_{n^\prime,m^\prime,n_{k^{c^{\prime}}}}(\tau)|^2 cos(n^\prime k.b+m^\prime k.c)}+\frac{\xi|W|^2}{{\cal V}^3},
\nonumber\\
& &
\end{eqnarray}
where ${\cal V}$ is the overall volume of the Swiss-Cheese Calabi-Yau, $n^s$ is the $D3$-brane instanton quantum number, $m^a$'s are the $D1$ instanton numbers and $f(\tau)$ is an appropriate modular function. The expressions for ${\cal Y}$, the holomorphic Jacobi theta function $\theta_{n^\alpha}(\tau,G)$ and $A_{n,m,n_{k^c}}(\tau)$ are defined as:

\begin{eqnarray}
\label{eq:nonpert14}
&& {\cal Y}\equiv {\cal V}_E + \frac{\chi}{2}\sum_{m,n\in{\bf Z}^2/(0,0)}
\frac{(\tau - {\bar\tau})^{\frac{3}{2}}}{(2i)^{\frac{3}{2}}|m+n\tau|^3} \nonumber\\
& &
- 4\sum_{\beta\in H_2^-(CY_3,{\bf Z})}n^0_\beta\sum_{m,n\in{\bf Z}^2/(0,0)}
\frac{(\tau - {\bar\tau})^{\frac{3}{2}}}{(2i)^{\frac{3}{2}}|m+n\tau|^3}cos\left((n+m\tau)k_a\frac{(G^a-{\bar G}^a)}{\tau - {\bar\tau}}
 - mk_aG^a\right),\nonumber\\
& & \theta_{n^\alpha}(\tau,G)=\sum_{m_a}e^{\frac{i\tau m^2}{2}}e^{in^\alpha G^am_a},\nonumber\\
& & A_{n,m,n_{k^c}}(\tau)\equiv \frac{(n+m\tau)n_{k^c}}{|n+m\tau|^3}.
\end{eqnarray}
Also, $G^a$ are defined by $G^a\equiv c^a-\tau b^a$ (where $c^a$'s and $b^a$'s are defined through the real RR two-form potential $C_2=c_a\omega^a$ and the real NS-NS two-form potential $B_2=b_a\omega^a$).

On comparing (\ref{eq:nonpert21}) with the analysis of \cite{Balaetal2}, one sees that for generic values of
the moduli $\rho_\alpha, G^a, k^{1,2}$ and ${\cal O}(1)\ W_{c.s.}$, and $n^s$(the $D3$-brane instanton quantum number)=1, analogous to \cite{Balaetal2}, the second term
dominates; the third term is a new term. However, as in KKLT scenarios (See \cite{KKLT}), $W_{c.s.}<<1$; we would henceforth assume that the fluxes and complex structure moduli have been so fine tuned/fixed that $W\sim W_{n.p.}$. We assume that the fundamental-domain-valued $b^a$'s satisfy: $\frac{|b^a|}{\pi}<1$\footnote{If one puts in appropriate powers of the Planck mass $M_p$, $\frac{|b^a|}{\pi}<1$ is equivalent to $|b^a|<\pi M_p$, i.e., NS-NS axions are sub-Planckian in units of $\pi M_p$.}. This implies that for $n^s>1$, the first term in (\ref{eq:nonpert21}) - $|\partial_{\rho^s}W_{np}|^2$ - a positive definite term, is the most dominant. In the same, $\rho^s$ is the volume of the small divisor complexified by RR 4-form axions. Hence, if a minimum exists, it will be positive. As shown in \cite{dSetal}, the potential can be extremized along the locus:
\begin{equation}
\label{eq:ext_locus}
mk.c + nk.b = N_{(m,n;,k^a)}\pi
\end{equation}
with $n^s>1$ and for all values of the $D1$-instanton quantum numbers $m^a$.\footnote{Considering the effect of axionic shift symmetry (of the $b^a$ axions) on the $D1$-instanton superpotential ($W_{D1-instanton}$), one can see that $m^a$ is valued in a lattice with coefficients being integral multiple of $2\pi$.}

As shown in section {\bf 3} of \cite{axionicswisscheese}, it turns out that the locus $nk.b + mk.c = N\pi$ for $|b^a|<\pi$ and $|c^a|<\pi$ corresponds to a flat saddle point with the NS-NS axions providing a flat direction.
For all directions in the moduli space with  $W_{c.s.}\sim {\cal O}(1)$ and away from $D_iW_{cs}=D_\tau W=0=\partial_{c^a}V=\partial_{b^a}V=0$, the ${\cal O}(\frac{1}{{\cal V}^2})$ contribution
of $\sum_{\alpha,{\bar\beta}\in{c.s.}}(G^{-1})^{\alpha{\bar\beta}}D_\alpha W_{cs}{\bar D}_{\bar\beta}{\bar W}_{cs}$  dominates over (\ref{eq:nonpert21}),
ensuring that that there must exist a minimum, and given the positive definiteness of the potential, this will be a dS minimum. There has been no need to add any $\overline{D3}$-branes as in KKLT to generate a dS vacuum.

\section{Finite $f_{NL}$}

We now proceed to showing the possibility of getting finite values for the non-linearity parameter $f_{NL}$ in two different contexts. First, we show the same for slow-roll inflationary scenarios. Second, we show the same when the slow-roll conditions are violated.

\subsection{Slow-Roll Inflationary Scenarios}

In \cite{axionicswisscheese}, we discussed the possibility of getting slow roll inflation along a flat direction provided by the NS-NS axions starting from a saddle point and proceeding to the nearest dS minimum. In what follows, we will assume that the volume moduli for the small and big divisors and the axion-dilaton modulus have been stabilized. All calculations henceforth will be in the axionic sector - $\partial_a$ will imply $\partial_{G^a}$ in the following.
On evaluation of the slow-roll inflation parameters (in $M_p=1$ units)
$\epsilon\equiv\frac{{\cal G}^{ij}\partial_iV\partial_jV}{2V^2},\ \eta\equiv$ the most negative eigenvalue of the matrix $N^i_{\ j}\equiv\frac{{\cal G}^{ik}\left(\partial_k\partial_jV - \Gamma^l_{jk}\partial_lV\right)}{V}$ with $\Gamma^l_{jk}$ being the affine connection components, we found that ${\epsilon}\sim{\frac{(n^s)^2}{(k^2 g_s^{\frac{3}{2}}\Delta){\cal V}}}$ and ${\eta}\sim{\frac{1}{k^2 g_s^{\frac{3}{2}} \Delta}}[g_s n^s \kappa_{1ab} +{\frac{(n^s)^2}{\sqrt{\cal V}}} \pm{n^s k^2 g_s^{\frac{3}{2}}\Delta}]$\footnote{The $g_s$ and k-dependence of $\epsilon$ and $\eta$ was missed in \cite{axionicswisscheese}. The point is that the extremization of the potential w.r.t.$b^a$'s and $c^a$'s in the large volume limit yields a saddle point at $sin(nk.b+mk.c)=0$ and those maximum degree-$k^a$ holomorphic curves $\beta$ for which $b^a\sim -m^a/{\kappa}$ (assuming that $\frac{nk.m}{\pi\kappa}\in{\bf Z}$).} where $\Delta\equiv{\frac{\sum_{\beta\in H_2^-(CY_3,{\bf Z})} n^{0}_{\beta}}{{\cal V}}}$ and we have chosen Calabi-Yau volume ${\cal V}$ to be such that ${\cal V}\sim e^{\frac{4\pi^2}{g_s}}$ (similar to \cite{LargeVcons}). Using Castelnuovo's theory of study of moduli spaces that are fibrations of Jacobian of curves over the moduli space of their deformations, for compact Calabi-Yau's expressed as projective varieties in weighted complex projective spaces (See \cite{Klemm_GV}) one sees that for appropriate degrees of the holomorphic curve, the genus-0 Gopakumar-Vafa invariants can be very large to compensate the volume factor appearing in the expression for $\eta$. Hence the slow-roll conditions can be satisfed, and in particular, there is no ``$\eta$"-problem. By investigating the eigenvalues of the Hessian, we showed (in \cite{axionicswisscheese}) that one could identify a linear combination of the NS-NS axions (``$k_2b^2+k_1b^1$") with the inflaton and the slow-roll inflation starts from the aforementioned saddle-point and ends when the slow-roll conditions were violated, which most probably corresponded to the nearest dS minimum, one can show that (in $M_p=1$ units)
\begin{equation}
\label{eq:Ne def}
N_e=-\int_{{\rm in:\ Saddle\ Point}}^{{\rm fin:\ dS\ Minimum}}\frac{1}{\sqrt{\epsilon}}d{\cal I}\sim
  \frac{k g_s^{3/4}\sqrt{\sum_{\beta\in H_2}n^0_\beta}}{n^s }.
\end{equation}  We will see that one can get $N_e\sim{60}$ e-foldings in the context of slow roll as well as slow roll violating scenarios.
Now before explaining how to get the non-linear parameter ``$f_{NL}$" relevant to studies of non-Gaussianities,
to be ${\cal O}(10^{-2})$ in our slow-roll LVS Swiss-Cheese orientifold setup, let us summarize  the formalism and results of \cite{Yokoyamafnl} in which the authors analyze the primordial non-Gaussianity in multi-scalar field inflation models (without the explicit form of the potential) using the slow-roll conditions and the $\delta N$ formalism of (\cite{Maldacenadeltan}) as the basic inputs.

Assuming that the time derivative of scalar field $\phi^a(t)$ is not independent of $\phi^a(t)$ (as in the case of standard slow-roll inflation) the background $e$-folding number between an initial hypersurface at $t=t_*$ and a final hypersurface at $t=t_{\rm c}$ (which is defined by
$N \equiv \int H dt$) can be regarded as a function of the homogeneous background field configurations $\phi^a(t_*)$ and $\phi^a(t_{\rm c})$ (on the initial and final hypersurface at $t=t_*$ and $t=t_{\rm c}$ respectively). i.e.
\begin{equation}
N\equiv N(\phi^a(t_{\rm c}),\phi^a(t_*))~.
\end{equation}
By considering $t_{\rm c}$ to be a time when the background trajectories have converged, the curvature perturbation $\zeta$ evaluated at $t=t_{\rm c}$ is given by $\delta N(t_{\rm c},\phi^a(t_*))$ (using the $\delta N$ formalism). After writing the $\delta N(t_{\rm c},\phi^a(t_*))$ upto second order in field perturbations $\delta{\phi^a(t_*)}$ (on the initial flat hypersurface at $t=t_*$) the curvature perturbation $\zeta(t_{\rm c})$ becomes
\begin{equation}
\label{deltaNsecond}
\zeta(t_{\rm c}) \simeq \delta N(t_{\rm c},\phi^a_{*}) = \partial_{a}N^*\delta
\phi^a_{*} + {1 \over 2}\partial_{a}\partial_{b}N^*\delta \phi^a_* \delta \phi^b_*~,
\end{equation}
and using the power spectrum correlator equations and $\zeta ({\bf x}) = \zeta_G({\bf x}) -{3 \over 5}f_{NL}\zeta_G^2({\bf x})$, where $\zeta_G({\bf x})$ represents the Gaussian part, one can arrive at
\begin{equation}
- {6 \over 5}f_{NL} \simeq{\partial^{a}N_*\partial^{b}N_*\partial_{a}\partial_{b}N^*\over
   \left(\partial_{c}N^*\partial^{c}N_*\right)^2}~
\label{fNLdeltaN}
\end{equation}
with the assumption that the field perturbation on the initial flat hypersurface, $\delta \phi^a_*$, is Gaussian.

For the generalization of the above in the context of non-Gaussianties, the authors assumed the so called ``relaxed" slow-roll conditions (RSRC) (which is $\epsilon \ll 1$ and $|\eta_{ab}| \ll 1$) for all the scalar fields, and introduce a time $t_{\rm f}$, at which the RSRC are still satisfied. Then for calculating  $\zeta(t_{\rm c})$, they express $\delta\phi^a(t_{\rm f})$ in terms of $\delta\phi^a_*$, with the scalar field expanded as $\phi^a \equiv {\phi_0}^a + \delta \phi^a$ and then evaluate  $N(t_{\rm c}, \phi^a(t_{\rm f}))$ (the $e$-folding number to reach $\phi^{a(0)}(t_{\rm c})$ starting with $\phi^a=\phi^a(t_{\rm f})$) and with the calculation of $\zeta(t_{\rm c})$ in terms of derivatives of field variations of $N^f$ (making use of the background field equations in variable $N$ instead of time variable) and comparing the same with (\ref{deltaNsecond}) and using (\ref{fNLdeltaN}) one arrives at the following general expression for the non-linear parameter $f_{NL}$:
\begin{equation}
\label{eq:fNL_def}
-\frac{6}{5}f_{NL} = \frac{\partial_a\partial_bN^f\Lambda^a_{a^\prime}{\cal G}^{a^\prime a^{\prime\prime}}\partial_{a^{\prime\prime}}N_* \Lambda^b_{b^\prime}{\cal G}^{b^\prime b^{\prime\prime}}\partial_{b^{\prime\prime}}N_* + \int_{N_*}^{N_f}dN \partial_cNQ^c_{df}\Lambda^d_{d^\prime}{\cal G}^{d^\prime d^{\prime\prime}}\partial_{d^{\prime\prime}}N_*
\Lambda^f_{f^\prime}{\cal G}^{f^\prime f^{\prime\prime}}\partial_{f^{\prime\prime}}N_*}{({\cal G}^{kl}\partial_kN^*\partial_lN_*)^2},
\end{equation}
with the following two constraints (See \cite{Yokoyamafnl}) required:
\begin{equation}
\label{eq:slowrollconstraint}
\biggl|\frac{{\cal G}^{ab}\left({\partial_b} V\right)_{;a}}{V}\biggr|\ll\sqrt{\frac{{\cal G}^{ab}\partial_{a} V\partial_{b}V}{V^2}}\ {\rm and}\
\biggl|Q^a_{bc}\biggr|\ll\sqrt{\frac{{\cal G}^{ab}\partial_{a} V\partial_{b}V}{V^2}},
\end{equation}
the semicolon implying a covariant derivative involving the affine connection.
In (\ref{eq:fNL_def}) and (\ref{eq:slowrollconstraint}), ${\cal G}^{ab}$'s are the components of the moduli space metric along the axionic directions given as,
\begin{equation}
\label{eq:inversemetric}
{\cal G}^{ab}\sim {\frac{{\cal Y}}{k^2 g_s^{\frac{3}{2}} \sum_{\beta\in H_2^-(CY_3,{\bf Z})} n^{0}_{\beta}}}\equiv{\frac{1}{k^2 g_s^{\frac{3}{2}} \Delta}}
\end{equation} Further,
\begin{eqnarray}
\label{eq:LambdaPQ_defs}
& & \Lambda^a_b\equiv \left(T e^{\int_{N_*}^{N_f}P(N)dN}\right)^a_b,\ P^a_b\equiv\left[-\frac{\partial_{a^\prime}({\cal G}^{aa^\prime}\partial_bV)}{V} + \frac{{\cal G}^{aa^\prime}\partial_{a^\prime}V\partial_bV}{V^2}\right];\nonumber\\
& & Q^a_{bc}\equiv\left[-\frac{\partial_{a^\prime}({\cal G}^{aa^\prime}\partial_b\partial_cV)}{V}
+\frac{\partial_{a^\prime}({\cal G}^{aa^\prime}\partial_{(b}V)\partial_{c)}V}{V^2}
+\frac{{\cal G}^{aa^\prime}\partial_{a^\prime}V\partial_b\partial_cV}{V^2}
-2\frac{{\cal G}^{aa^\prime}\partial_{a^\prime}V\partial_bV\partial_cV}{V^3}\right],
\end{eqnarray}
where $V$ is the scalar potential and as the number of e-folding is taken as a measure of the period of inflation (and hence as the time variable), the expression for $\Lambda^I_J$ above, has a time ordering $T$ with the initial and final values of number of e-foldings $N_*$ and $N_f$ respectively. From the definition of $P^I_J$ and $\Lambda^I_J$, one sees that during the slow-roll epoch, $\Lambda^I_J=\delta^I_J$.

After using (\ref{eq:nonpert21}) along with:
\begin{eqnarray}
\label{eq:constraints}
& & \sum_{m_a\in 2{\bf Z}\pi}{e^{-\frac{m^2}{2g_s} + \frac{m_ab^a n^s}{g_s} + \frac{n^s\kappa_{1ab}b^ab^b}{2g_s}}}\sim 1 \nonumber\\
& & \sum_{m_a\in 2{\bf Z}\pi} m_a {e^{-\frac{m^2}{2g_s} + \frac{m_ab^a n^s}{g_s} + \frac{n^s\kappa_{1ab}b^ab^b}{2g_s}}}\sim e^{\frac{-2\pi^2}{g_s}}\sim \frac{1}{\sqrt{\cal V}}, \nonumber\\
& & \end{eqnarray}
for sub-planckian $b^a$'s, one arrives at the following results (along the slow-roll direction $sin(n k_ab^a+m k_ac^a)=0$) :
\begin{eqnarray}
\label{eq:derpotential}
& & \frac{\partial_{a} V}{V}\sim \frac{n^s}{\sqrt{\cal V}};
\frac{\partial_{a}\partial_{b} V}{V}\sim g_s n^s \kappa_{1ab} +\frac{(n^s)^2}{\sqrt{\cal V}} \pm n^s k^2 g_s^{\frac{7}{2}}\Delta,\nonumber\\
& & \frac{\partial_a \partial_{b}\partial_{c}V}{V} \sim {\frac{n^s}{\sqrt{\cal V}}}\Biggl[g_s (n^s) \kappa_{1ab} +(n^s)^2  \pm g_s^{\frac{7}{2}} n^s k^2 \Delta\Biggr]
\end{eqnarray}
Further using the above, one sees that the $\epsilon$ and $\eta$ parameters along with $Q^a_{bc}$ (appearing in the expression of $f_{NL}$) are given as under:
\begin{equation}
\label{eq:epeta}
\epsilon\sim {\frac{(n^s)^2}{{\cal V} g_s^{\frac{3}{2}} k^2 \Delta}}; \eta \sim \frac{1}{g_s^{\frac{3}{2}}k^2 \Delta}\Biggl[ g_s n^s \kappa_{1ab} +\frac{(n^s)^2}{\sqrt{\cal V}} \pm {n^s k^2 g_s^{\frac{7}{2}} \Delta}\Biggr],
\end{equation}
and
\begin{equation}
\label{eq:Q}
Q^a_{bc}\sim \frac{n^s}{{g_s}^{\frac{3}{2}}k^2\Delta{\sqrt{\cal V}}}\biggl[(n^s)^2 +\frac{(n^s)^2}{\sqrt{\cal V}}-\frac{(n^s)^2}{{\cal V}}\pm {n^s k^2 g_s^{\frac{7}{2}}\Delta}\biggr]
\end{equation}
Now in order to use the expression for $f_{NL}$, the first one of required constraints  (\ref{eq:slowrollconstraint}) results in the following inequality:
\begin{eqnarray}
\label{eq:srconstraint1}
|\delta|\equiv\Biggl|g_s n^s \kappa_{1ab} +\frac{(n^s)^2}{\sqrt{\cal V}} - n^s g_s^{\frac{7}{2}} k^2 \Delta\Biggr|\ll{n^s}{\sqrt\frac{g_s^{\frac{3}{2}} k^2 \Delta}{{\cal V}}}
\end{eqnarray}
Now we solve the above inequality for say $|\delta|\sim{\frac{1}{{\cal V}}}$, which is consistent with the constraint requirement along with the following relation
\begin{equation}
\label{eq:Delta}
\Delta\equiv\frac{\sum_{\beta\in H_2^-(CY_3,{\bf Z})} n^{0}_{\beta}}{\cal Y}\sim {\frac{1}{k^2 g_s^{\frac{7}{2}}}\Biggl[g_s \kappa_{1ab}+\frac{n^s}{\sqrt{\cal V}}}\Biggr]\sim{\frac{1}{k^2 g_s^{\frac{5}{2}}}}
\end{equation}
The second constraint is
\begin{equation}
\label{eq:constraint_2}
\Biggl| (n^s)^2+\frac{(n^s)^2}{\sqrt{\cal V}}+\frac{(n^s)^2}{{\cal V}}- g_s^{\frac{7}{2}} k^2 (n^s)\Delta\Biggr|\ll{\sqrt{k^2 \Delta g_s^{\frac{3}{2}}}}
\end{equation}
Given that we are not bothering about precise numerical factors, we will be happy with ``$<$" instead of a strict ``$\ll$" in (\ref{eq:constraint_2}).
Using (\ref{eq:Delta}) in the previous expressions (\ref{eq:epeta},\ref{eq:Q}) for $\epsilon,\eta$ and $Q^a_{bc}$, we arrive at the final expression for the slow-roll parameters $\epsilon,\eta$ and $Q^a_{bc}$ as following:
\begin{eqnarray}
& & \epsilon \sim {G^{ab}}\frac{(n^s)^2}{{\cal V}}\sim \frac{g_s (n^s)^2}{{\cal V}}; |\eta| \sim G^{ab}|\delta|\sim\frac{g_s }{{\cal V}}\nonumber\\
& & (Q^a_{bc})_{max}\sim G^{ab}{\sqrt{k^2 \Delta g_s^{\frac{3}{2}}}}\biggl({\frac{n^s}{{\sqrt{\cal V}}}}\biggr)\sim{n^s{\sqrt\frac{g_s}{{\cal V}}}}
\end{eqnarray}

As the number of e-foldings satisfies $\partial_{I}N=\frac{V}{\partial_I V}\sim\frac{\sqrt{{\cal V}}}{n^s \sqrt{g_s}}$, which is almost constant and hence $\partial_{I}\partial_{J}N\sim0$. Consequently the first term of (\ref{eq:fNL_def}) is negligible and the maximum contribution to the non-Gaussianities parameter $f_{NL}$ coming from the second term is given by:

\begin{equation}
\label{eq:fNLII}
\frac{\int_{N_*}^{N_f}dN \partial_cNQ^a_{bc}\Lambda^b_{b^\prime}{\cal G}^{b^\prime b^{\prime\prime}}\partial_{b^{\prime\prime}}N
\Lambda^c_{c^\prime}{\cal G}^{c^\prime c^{\prime\prime}}\partial_{c^{\prime\prime}}N}{({\cal G}^{df}\partial_dN\partial_fN)^2}\le(Q^a_{bc})_{max}\sim{n^s{\sqrt\frac{g_s}{{\cal V}}}}.
\end{equation}

This way, for Calabi-Yau volume ${\cal V}\sim{10}^{6}$, D3-instanton number $n^s={\cal O}(10)$ with $n^s\sim g_s\sim k^2$implying the slow-roll parameters\footnote{These values are allowed for the curvature perturbation freeze-out at the superhorizon scales, which is discussed in the section pertaining to finite tensor-to scalar ratio.} $\epsilon\sim{0.00028}, |\eta|\sim {10}^{-6}$ with the number of e-foldings $N_e\sim{60}$, one obtains the maximum value of the non-Gaussianties parameter $\left(f_{NL}\right)_{\rm max}\sim {10}^{-2}$. Further if we choose the stabilized Calabi-Yau volume ${\cal V}\sim{10}^{5}$ with $n^s={\cal O}(10)$, we find $\epsilon\sim{0.0034}, |\eta|\sim {10}^{-4}$ with the number of e-foldings $N_e\sim{17}$ and the maximum possible $\left(f_{NL}\right)_{\rm max}\sim 3\times{10}^{-2}$. The above mentioned values of $\epsilon$ and $\eta$ parameters can be easily realized in our setup with the appropriate choice of
holomorphic isometric involution as part of the Swiss-Cheese orientifold. This way we have realized ${\cal O}({10}^{-2})$ non-gaussianities parameter $f_{NL}$ in slow-roll scenarios of our LVS Swiss-Cheese orientifold setup.

\subsection{Slow-Roll Conditions are Violated}

We will now show that it is possible to obtain ${\cal O}(1)$ $f_{NL}$ while looking for non-Gaussianities in curvature perturbations when the slow-roll conditions are violated.
We will follow the formalism developed in \cite{Yokoyama} to discuss evaluation of $f_{NL}$ in scenarios wherein the slow-roll conditions are violated. Before that let us summarize the results of \cite{Yokoyama} in which the authors analyze the non-Gaussianity of the primordial curvature perturbation generated on super-horizon scales in multi-scalar field inflation models {\it without imposing the slow-roll conditions} and using the $\delta N$ formalism of (\cite{Maldacenadeltan}) as the basic input.

Consider a model with $n-$component scalar field $\phi^a$.  Now consider the perturbations of the scalar fields in constant $N$ gauge as
\begin{eqnarray}
\delta \phi^{\cal A} (N)  \equiv \phi^{\cal A}( \lambda+\delta\lambda; N)-\phi^{\cal A}(\lambda; N),
\end{eqnarray}
where the short-hand notation of \cite{Yokoyama} is used - $X^{\cal A}\equiv X^a_{i}(i=1,2)=(X^a_1\equiv X^a,X^a_2\equiv\frac{dX^a}{dN})$\footnote{We have modified the notations of \cite{Yokoyama} a little.}, and
where $\lambda^{\cal A}$'s are the $2n$ integral constants of the background field equations. After using the decomposition of the fields $\phi^{\cal A}$ up to second order in $\delta$ (defined through $\delta \tilde{\phi}^{\cal A}=\delta \tilde{\phi}_{(1)}^{\cal A}+{1 \over 2}\delta \tilde{\phi}_{(2)}^{\cal A}$; to preserve covariance under general coordinate transformation in the moduli space, the authors of \cite{Yokoyama} define: $(\delta \tilde{\phi}_{(1)})^a_1\equiv\frac{d\phi^a}{d\lambda}\delta\lambda$,\
$(\delta \tilde{\phi}_{(2)})^a_1\equiv\frac{D}{d\lambda}\frac{d\phi^a}{d\lambda}(\delta\lambda)^2$ and
$(\delta \tilde{\phi}_{(1)})^a_2\equiv\frac{D\phi^a_2}{d\lambda}\delta\lambda$, $(\delta \tilde{\phi}_{(2)})^a_2\equiv\frac{D^2\phi^a_2}{d\lambda^2}(\delta\lambda)^2$), one can solve the evolution equations for $\delta \tilde{\phi}_{(1)}^{\cal A}$ and $\delta \tilde{\phi}_{(2)}^{\cal A}$. The equation for $\delta \tilde{\phi}_{(2)}^{\cal A}$ is simplified with the choice of integral constants such that $\lambda^{\cal A}=\phi^{\cal A}(N_*)$ implying $\delta \tilde{\phi}^{\cal A}(N_*)=\delta \lambda^{\cal A}$ and hence $\delta\tilde{\phi}_{(2)}^{\cal A}(N)$ vanishing at $N_*$. Assuming $N_*$ to be a certain time soon after the relevant length scale crossed the horizon scale ($H^{-1}$), during the scalar dominant phase and $N_c$ to be a certain time after the complete convergence of the background trajectories has occurred and using the so called $\delta N$ formalism one gets
\begin{equation}
\zeta \simeq\delta N = \tilde{N}_{{\cal A}}\delta \tilde{\phi}^{\cal A} + {1 \over 2}\tilde{N}_{{\cal A}{\cal B}}\delta \tilde{\phi}^{\cal A} \delta \tilde{\phi}^{\cal B} +\cdots
\end{equation}

Now taking $N_f$ to be certain late time during the scalar dominant phase and using the solutions for $\delta \phi_{(1)}^{\cal A}$ and $\delta \phi_{(2)}^{\cal A}$ for the period $N_*<N<N_f$, one obtains the expressions for $\tilde{N}_{{\cal A}*}$ and $\tilde{N}_{{\cal A}{\cal B}*}$ (to be defined below) and finally writing the variance of $\delta \tilde{\phi}^{\cal A}_*$ (defined through
$\langle \delta \tilde{\phi}^{\cal A}_* \delta \tilde{\phi}^{\cal B}_*\rangle
\simeq A^{{\cal A}{\cal B}}{\left( \frac{H_\ast}{2\pi} \right)}^2$ including  corrections to the slow-roll terms in $A^{ab}$ based on \cite{Byrnes_Wands,Byrnestrispectrum}), and using the basic definition of the non- linear parameter $f_{NL}$ as the the magnitude of the bispectrum of the curvature perturbation $\zeta$, one arrives at a general expression for $f_{NL}$ (for non slow roll cases)\cite{Yokoyama}. For our present interest, the expression for $f_{NL}$ for the non-slow roll case is given by
\begin{equation}
\label{eq:fNL_defii}
-\frac{6}{5}f_{NL} = \frac{\tilde{N}^f_{{\cal A}{\cal B}}\Lambda^{\cal A}_{{\cal A}^\prime}(N_f,N_*)A^{{\cal A}^\prime {\cal A}^{\prime\prime}}\tilde{N}^*_{{\cal A}^{\prime\prime}} \Lambda^{\cal B}_{{\cal B}^\prime}A^{{\cal B}^\prime {\cal B}^{\prime\prime}}\tilde{N}^*_{{\cal B}^{\prime\prime}} + \int_{N_*}^{N_f}dN \tilde{N}_{\cal C}\tilde{Q}^{\cal C}_{{\cal D}{\cal F}}\Lambda^{\cal D}_{{\cal D}^\prime}A^{{\cal D}^\prime {\cal D}^{\prime\prime}}\tilde{N}^*_{{\cal D}^{\prime\prime}}
\Lambda^{\cal F}_{{\cal F}^\prime}A^{{\cal F}^\prime {\cal F}^{\prime\prime}}\tilde{N}^*_{{\cal F}^{\prime\prime}}}{(A^{{\cal K}{\cal L}}\tilde{N}^*_{\cal K}\tilde{N}^*_{\cal L})^2},
\end{equation}
where again the index ${\cal A}$ represents a pair of indices $^a_i$, $i=1$ corresponding to the field $b^a$ and $i=2$ corresponding to $\frac{db^a}{dN}$. Further,
\begin{eqnarray}
\label{eq:Adef}
& & A^{ab}_{11}={\cal G}^{ab}+\left(\sum_{m_1,m_2,m_3,m_4,m_5}^{<\infty}\left(||\frac{d\phi^a}{dN}||^2\right)^{m_1}
\left(\frac{1}{H}\frac{dH}{dN}\right)^{m_2}\epsilon^{m_3}\eta^{m_4}\right)^{ab},\nonumber\\
& & A^{ab}_{12}=A^{ab}_{21}=\frac{{\cal G}^{aa^\prime}\partial_{a^\prime}V{\cal G}^{bb^\prime}\partial_{b^\prime}}{V^2}-\frac{\partial_{a^\prime}({\cal G}^{aa^\prime}{\cal G}^{bb^\prime}\partial_{b^\prime}V)}{V},\nonumber\\
& & A^{ab}_{22}=\left(\frac{{\cal G}^{aa^\prime}\partial_{a^\prime}V\partial_cV}{V^2}-\frac{\partial_{a^\prime}({\cal G}^{aa^\prime}\partial_cV)}{V}\right)\left(\frac{{\cal G}^{cc^\prime}\partial_{c^\prime}V{\cal G}^{bb^\prime}\partial_{b^\prime}V}{V^2}-\frac{\partial_{c^\prime}({\cal G}^{cc^\prime}{\cal G}^{bb^\prime}\partial_{b^\prime}V)}{V}\right),
\end{eqnarray}
where in $A^{ab}_{11}$, based on \cite{Byrnes_Wands,Byrnestrispectrum}, assuming the non-Gaussianity to be expressible as a finite-degree polynomial in higher order slow-roll parameter corrections.
In (\ref{eq:fNL_defii}), one defines:
\begin{equation}
\label{eq:Lambda}
\Lambda^{\cal A}_{\ {\cal B}}=\left(Te^{\int_{N_*}^{N_f}dN \tilde{P}(N)}\right)^{\cal A}_{\cal B};
\end{equation}
$\tilde{N}_{\cal A}, \tilde{N}_{{\cal A}{\cal B}},\tilde{P}^{\cal A}_{\cal B}$ and $\tilde{Q}^{\cal A}_{\ {\cal B}{\cal C}}$\footnote{We have modified the notations of \cite{Yokoyama} for purposes of simplification.} will be defined momentarily. The equations of motion
\begin{eqnarray}
\label{eq:eoms}
& & \frac{d^2b^a}{dN^2} + \Gamma^a_{bc}\frac{db^b}{dN}\frac{db^c}{dN} + \left(3+\frac{1}{H}\frac{dH}{dN}\right)\frac{db^a}{dN}+\frac{{\cal G}^{ab}\partial_bV}{H^2}=0,\nonumber\\
& & H^2=\frac{1}{3}\left(\frac{1}{2}H^2||\frac{db^a}{dN}||^2 + V\right)
\end{eqnarray}
yield
\begin{equation}
\label{eq:conseqEOM}
\frac{1}{H}\frac{dH}{dN}=\frac{6\left(\frac{2V}{H^2}-6\right) - 2\Gamma^a_{bc}\frac{db_a}{dN}\frac{db^b}{dN}\frac{db^c}{dN}}{12}.
\end{equation}

For slow-roll inflation, $H^2\sim\frac{V}{3}$; the Friedmann equation in (\ref{eq:eoms}) implies that
$H^2>\frac{V}{3}$ when slow-roll conditions are violated. The number of e-foldings away from slow-roll
is given by: $N\sim\int\frac{db^a}{||\frac{db^b}{dN}||}$, which using the Friedmann equation implies
$N\sim\int\frac{db^a}{\sqrt{1-\frac{V}{3H^2}}}$.

We would require $\epsilon<<1$ and $|\eta|~{\cal O}(1)$ to correspond to a slow-roll violating scenario. Now, writing the potential $V=V_0\frac{\sqrt{ln {\cal Y}}}{{\cal Y}^{2n^s+1}}\left(\sum_{m\in2{\bf Z}\pi}e^{-\frac{m^2}{2g_s}+\frac{n^s}{g_s}m.b+\frac{\kappa_{1ab}b^ab^b}{2g_s}}\right)^2$, $\partial_{{\cal G}^a}V=0$ implies:
\begin{eqnarray}
\label{eq:slow_roll_viol_I}
& & \partial_{{\cal G}^a}V\sim-\frac{g_s\sqrt{ln {\cal Y}}n^s}{{\cal Y}^{2n^s+1}}\sum_{\beta\in H_2^-(CY_3,{\bf Z})}\frac{n^0_\beta}{{\cal Y}}sin(nk.b+mk.c)k^ag_s^{\frac{3}{2}}\left(\sum_{m\in2{\bf Z}\pi}e^{-\frac{m^2}{2g_s}+\frac{n^s}{g_s}m.b+\frac{\kappa_{1ab}b^ab^b}{2g_s}}\right)^2 \nonumber\\
& & +
\frac{g_s\sqrt{ln {\cal Y}}}{{\cal Y}^{2n^s+1}}\left(\sum_{m\in2{\bf Z}\pi}e^{-\frac{m^2}{2g_s}+\frac{n^s}{g_s}m.b+\frac{\kappa_{1ab}b^ab^b}{2g_s}}\right)
\frac{n^s}{g_s}\sum_{m\in2{\bf Z}\pi}(m^a+\kappa_{1ab}b^b)\left(\sum_{m^a\in2{\bf Z}\pi}e^{-\frac{m^2}{2g_s}+\frac{n^s}{g_s}m.b+\frac{\kappa_{1ab}b^ab^b}{2g_s}}\right),
\end{eqnarray}
which using $\sum_{m\in2{\bf Z}\pi}e^{-\frac{m^2}{2g_s}+\frac{n^s}{g_s}m.b+\frac{\kappa_{1ab}b^ab^b}{2g_s}}\sim e^{\frac{\kappa_{1ab}b^ab^b}{2g_s}}$ and $\sum_{m\in2{\bf Z}\pi}(m^a+\kappa_{1ab}b^b)\left(\sum_{m^a\in2{\bf Z}\pi}e^{-\frac{m^2}{2g_s}+\frac{n^s}{g_s}m.b+\frac{\kappa_{1ab}b^ab^b}{2g_s}}\right)\sim\kappa_{1ab}b^b
e^{\frac{\kappa_{1ab}b^ab^b}{2g_s}}$, one obtains:
\begin{equation}
\label{eq:slow_roll_viol_II}
\sum_{\beta\in H_2^-(CY_3,{\bf Z})}\frac{n^0_\beta}{{\cal Y}}sin(nk.b+mk.c)k^a\sim\frac{b^a}{g_s^{\frac{5}{2}}}.
\end{equation}
Slow-roll scenarios assumed that the LHS and RHS of (\ref{eq:slow_roll_viol_II}) vanished individually - the same will not be true for slow-roll violating scenarios. Near (\ref{eq:slow_roll_viol_II}), one can argue that:
\begin{eqnarray}
\label{eq:Ginv_Gamma_I}
& & {\cal G}^{{\cal G}^a{\bar{\cal G}}^b}\sim\frac{b^ab^b - \sqrt{g_s^7\left(k^ak^b\right)^2\left(\frac{n^0_\beta}{\cal V}\right)^2 - b^2g_s^2k^2}}{b^2\sqrt{g_s^7\left(k^ak^b\right)^2\left(\frac{n^0_\beta}{\cal V}\right)^2 - b^2g_s^2k^2} + g_s^7\left(k^ak^b\right)^2\left(\frac{n^0_\beta}{\cal V}\right)^2 - b^2g_s^2k^2};\nonumber\\
& & \Gamma^{{\cal G}^a}_{{\cal G}^b{\cal G}^c}\sim\frac{\left(\sqrt{g_s^7\left(k^ak^b\right)^2\left(\frac{n^0_\beta}{\cal V}\right)^2 - b^2g_s^2k^2} + g_s^2 + \frac{b^2g_s^2}{\sqrt{g_s^7\left(k^ak^b\right)^2\left(\frac{n^0_\beta}{\cal V}\right)^2 - b^2g_s^2k^2}}\right)}{b^2\sqrt{g_s^7\left(k^ak^b\right)^2\left(\frac{n^0_\beta}{\cal V}\right)^2 - b^2g_s^2k^2} + g_s^7\left(k^ak^b\right)^2\left(\frac{n^0_\beta}{\cal V}\right)^2 - b^2g_s^2k^2}.
\end{eqnarray}
Note, we no longer restrict ourselves to sub-Planckian axions - we only require $|b^a|<\pi$. For $g_s^7\left(k^ak^b\right)^2\left(\frac{n^0_\beta}{\cal V}\right)^2 - b^2g_s^2k^2\sim {\cal O}(1)$, and the holomorphic isometric involution, part of the Swiss-Cheese orientifolding, assumed to be such that the maximum degree of the holomorphic curve being summed over in the non-perturbative $\alpha^\prime$-corrections involving the genus-zero Gopakumar-Vafa invariants are such that $\sum_\beta\frac{n^0_\beta}{\cal V}\leq\frac{1}{60}, k\sim 3$, we see that (\ref{eq:slow_roll_viol_II}) is satisfied and
\begin{eqnarray}
\label{eq:Ginv_Gamma_II}
& & {\cal G}^{{\cal G}^a{\bar{\cal G}}^b}\sim\frac{b^2+{\cal O}(1)}{b^2+{\cal O}(1)}\sim{\cal O}(1);\nonumber\\
& & \Gamma^{{\cal G}^a}_{{\cal G}^b{\cal G}^c}\sim\frac{b^a\left(g_s^2+b^2g_s^2\right)}{b^2+{\cal O}(1)}\sim b^ag_s^2.
\end{eqnarray}
Hence, the affine connection components, for $b^a\sim {\cal O}(1)$, is of ${\cal O}(10)$; the curvature components $R^a_{\ bcd}$ will hence also be finite. Assuming $H^2\sim V$, the definitions of $\epsilon$ and $\eta$ continue to remain the same as those for slow-roll scenarios and one hence obtains:
\begin{eqnarray}
\label{eq:epsilon_eta_slow_roll_viol}
& & \epsilon\sim\frac{\left(n^s\right)^2e^{\frac{4\pi n^sb}{g_s}+\frac{b^2}{g_s}}\left(\pi+b\right)^2}{\cal V}\sim10^{-3},\nonumber\\
& & \eta\sim n^s(1+n^sb^2)-\frac{bg_s^2n^se^{\frac{4\pi n^sb}{g_s}+\frac{b^2}{g_s}}\left(\pi+b\right)}{\sqrt{\cal V}}\sim n^s(1+n^sb)\sim {\cal O}(1).
\end{eqnarray}
Finally, $\frac{db^a}{dN}\sim\sqrt{1-\frac{V}{3H^2}}\sim{\cal O}(1)$.

We now write out the various components of $\tilde{P}^{\cal A}_{\cal B}$, relevant to evaluation of
$\Lambda^{\cal A}_{\ {\cal B}}$ in (\ref{eq:Lambda}):
\begin{eqnarray}
\label{eq:P's}
& & \tilde{P}^{a1}_{1b}=0,\nonumber\\
& & \tilde{P}^{a2}_{1b}=\delta^a_b,\nonumber\\
& & \tilde{P}^{a1}_{2b}=-\frac{V}{H^2}\left(\frac{\partial_a({\cal G}^{ac}\partial_cV)}{V}
- \frac{{\cal G}^{ac}\partial_cV\partial_bV}{V^2}\right) - R^a_{\ bcd}\frac{db^c}{dN}\frac{db^d}{dN}\sim{\cal O}(1),\nonumber\\
& & \tilde{P}^{a2}_{b2}={\cal G}_{bc}\frac{db^a}{dN}\frac{db^c}{dN}+\frac{{\cal G}^{ac}\partial_cV}{V}{\cal G}_{bf}\frac{db^f}{dN}-\frac{V}{H^2}\delta^a_b\sim{\cal O}(1).
\end{eqnarray}
Similarly,
\begin{eqnarray}
\label{eq:N's}
(a) & & N^1_a\sim\frac{1}{||\frac{db^a}{dN}||}\sim{\cal O}(1),\nonumber\\
& & N^2_a=0;\nonumber\\
(b) & & N^{11}_{ab}=N^{12}_{ab}=N^{22}_{ab}=0,\nonumber\\
& & N^{21}_{ab}\sim\frac{{\cal G}_{bc}\frac{db^c}{dN}}{||\frac{db^d}{dN}||^3}\sim{\cal O}(1);\nonumber\\
(c) & & \tilde{N}^1_a\equiv N^1_a-N^2_b\Gamma^b_{ca}\frac{db^c}{dN}\sim\frac{1}{||\frac{db^a}{dN}||}\sim{\cal O}(1),\nonumber\\
& & \tilde{N}^2_a\equiv N^2_a=0;\nonumber\\
(d) & & \tilde{N}^{11}_{ab}\equiv N^{11}_{ab}+N^{22}_{ca}\Gamma^c_{ml}\Gamma^l_{nb}\frac{db^m}{dN}\frac{db^n}{dN}+(N^{12}_{ac} + N^{21}_{ac})\Gamma^c_{lb}\frac{db^l}{dN}
-N^2_c(\bigtriangledown_a\Gamma^c_{lb})\frac{db^l}{dN}-N^2_c\Gamma^c_{al}\Gamma^l_{nb}\frac{db^n}{dN}\nonumber\\
& & \sim\frac{{\cal G}_{cd}\frac{db^d}{dN}}{||\frac{db^m}{dN}||^3}\Gamma^c_{lb}\frac{db^l}{dN}
\sim{\cal O}(1),\nonumber\\
& & \tilde{N}^{12}_{ab}\equiv N^{12}_{ab} - N^2_c\Gamma^c_{ab}-N^{22}_{cb}\Gamma^c_{al}\frac{db^l}{dN}=0,\nonumber\\
& & \tilde{N}^{21}_{ab}\equiv N^{21}_{ab}-N^2_c\Gamma^c_{ab}-N^{22}_{ca}\Gamma^c_{bl}\frac{db^l}{dN}\sim\frac{{\cal G}_{bc}\frac{db^c}{dN}}{||\frac{db^m}{dN}||^3}\sim{\cal O}(1),\nonumber\\
& & \tilde{N}^{22}_{ab}\equiv N^{22}_{ab}=0.
\end{eqnarray}
Finally,
\begin{eqnarray}
\label{eq:Q's}
& & \tilde{Q}^{a11}_{1bc}=-R^a_{\ bcd}\frac{db^d}{dN}\sim{\cal O}(1),\nonumber\\
& & \tilde{Q}^{a21}_{1bc}=\tilde{Q}^{a12}_{1bc}=\tilde{Q}^{a22}_{1bc}=0,\nonumber\\
& & \tilde{Q}^{a12}_{2bc}=\frac{\partial_a({\cal G}^{ad}\partial_dV)}{V}\frac{db^l}{dN}{\cal G}_{lc}-2R^a_{\ cbl}\frac{db^d}{dN}\sim{\cal O}(1),\nonumber\\
& & \tilde{Q}^{a22}_{2bc}=\delta^a_c{\cal G}_{bd}\frac{db^d}{dN}+\delta^a_b{\cal G}_{cd}\frac{db^d}{dN}
+{\cal G}_{bc}\left(\frac{db^a}{dN}+\frac{{\cal G}^{ad}\partial_aV}{V}\right)\sim{\cal O}(1),\nonumber\\
& & \tilde{Q}^{a11}_{2bc}=-\frac{V}{H^2}\left(\frac{\partial_a\partial_b({\cal G}^{ad}\partial_dV)}{V}-\frac{\partial_b({\cal G}^{ad}\partial_dV)\partial_cV}{V^2}\right)
- (\bigtriangledown_cR^a_{\ mbl})\frac{db^m}{dN}\frac{db^l}{dN}\sim{\cal O}(1),\nonumber\\
& & \tilde{Q}^{a21}_{2bc}=\left(\frac{\partial_c({\cal G}^{df}\partial_fV)}{V}-\frac{{\cal G}^{ad}\partial_dV\partial_cV}{V^2}\right){\cal G}_{bd}\frac{db^d}{dN}-R^a_{\ lcb}\frac{db^l}{dN}\sim{\cal O}(1).
\end{eqnarray}
So, substituting (\ref{eq:Adef}), (\ref{eq:N's})-(\ref{eq:Q's}) into (\ref{eq:fNL_defii}), one sees that $f_{NL}\sim{\cal O}(1)$.

After completion of this work, we were informed about \cite{largefNLloop} wherein observable values of $f_{NL}$ may be obtained by considering loop corrections.

\section{Finite Tensor-Scalar Ratio and Loss of Scale Invariance}

We now turn to looking for ``finite" values of ratio of ampltidues of tensor and scalar perturbations, ``$r$". Using the Hamilton-Jacobi formalism (See \cite{beyondslowroll} and references therein), which is suited to deal with beyond slow-roll approximations as well, the mode $u_k(y)$ - $y\equiv\frac{k}{aH}$ - corresponding to scalar perturbations, satisfies the following differential equation when one {\it does not assume slow roll conditions in the sense that even though $\epsilon$ and $\eta$ are still constants, but $\epsilon$ though less than unity need not be much smaller than unity and $|\eta|$ can even be of ${\cal O}(1)$ (See \cite{beyondslowroll})}\footnote{In this section, unlike subsection {\bf 3.2}, to simplify calculations, we would be assuming that one continues to remain close to the locus $sin(nk.b+mk.c)=0$ implying that the axionic moduli space metric is approximately a constant and the axionic kinectic terms, and in particular the inflaton kinetic term, with a proper choice of basis - see \cite{axionicswisscheese} - can be cast into a diagonal form. The cases having to do with being away from the slow-roll scenarios are effected by an appropriate choice of the holomorphic isometric involution involved in the Swiss-Cheese Calabi-Yau orientifold and (\ref{eq:slow_roll_viol_II}) } :
\begin{equation}
\label{eq:scalar1}
y^2(1-\epsilon)^2u^{\prime\prime}_k(y) + 2y\epsilon(\epsilon-\tilde{\eta})u_k^\prime(y)+\left(y^2 - 2\left(1+\epsilon-\frac{3}{2}\tilde{\eta}+\epsilon^2-2\epsilon\tilde{\eta}+\frac{\tilde{\eta}^2}{2}+\frac{\xi^2}{2}\right)\right)u_k(y)=0.
\end{equation}
In this section, following \cite{beyondslowroll}, we would be working with $\tilde{\eta}\equiv\eta-\epsilon$ instead of $\eta$.
We will be assuming that the slow parameter $\xi<<<1$\footnote{From \cite{beyondslowroll}, $\xi^2=\epsilon\tilde{\eta}-\epsilon\frac{d\tilde{\eta}}{d{\cal I}}$, ${\cal I}$ being the inflaton of section {\bf 3.1 }. Neglecting $\xi$ can be effected if $\tilde{\eta}\sim e^{\int\sqrt{\epsilon}d{\cal I}}$ - this is hence appropriate for hybrid inflationary scenarios, which is what gets picked out to ensure that the curvature perturbations do not grow at horizon crossing and beyond.} - this can be easily relaxed. In order to get a the required Minkowskian free-field solution in the long wavelength limit - the following is the solution\footnote{We follow \cite{Linde_Book} and hence choose $H^{(2)}_\nu(\frac{y}{-1+\epsilon})$ as opposed to $H^{(1)}_\nu(-\frac{y}{-1+\epsilon})$}:
\begin{equation}
\label{eq:scalar2}
u_k(y)\sim c(k) y^{\frac{1-\epsilon^2+2\epsilon(-1+\tilde{\eta})}{2(-1+\epsilon)^2}}H^{(2)}_{\frac{\sqrt{9+9\epsilon^4-12\tilde{\eta}+4\tilde{\eta}^2-4\epsilon^3(1+5\tilde{\eta})
-4\epsilon(3-3\tilde{\eta}+2\tilde{\eta}^2)+2\epsilon^2(1+6\tilde{\eta}+4\tilde{\eta}^2)}}{2(-1+\epsilon)^2}}
\left(\frac{y}{(-1+\epsilon)}\right).
\end{equation}
Now, $H^{(2)}_\alpha\equiv J_\alpha-i\left(\frac{J_\alpha cos(\alpha\pi) - J_{-\alpha}}{sin(\alpha\pi)}\right)$. \footnote{One would be interested in taking the small-argument limit of the Bessel function. However, the condition for doing the same, namely $0<\left|\frac{y}{-1+\epsilon}\right|<<\sqrt{\nu+1}$ is never really satisfied. One can analytically continue the Bessel function by using the fact that   $J_{\tilde{\tilde{\nu}}}(\frac{y}{(-1+\epsilon)})$ can be related to the Hypergeometric function $\ _0F_1\left(\tilde{\tilde{\nu}}+1;-\frac{y^2}{4(1-\epsilon)^2}\right)$ as follows:
\begin{equation}
\label{eq:J0F1}
J_{\tilde{\tilde{\nu}}}(\frac{y}{(1-\epsilon)})=\frac{\left(\frac{y}{2(1-\epsilon)}\right)^{\tilde{\tilde{\nu}}}}
{\Gamma(\tilde{\tilde{\nu}}+1)}\ _0F_1\left(\tilde{\tilde{\nu}}+1;-\frac{y^2}{4(1-\epsilon)^2}\right).
\end{equation}
where $\tilde{\tilde{\nu}}\equiv \frac{\sqrt{9+9\epsilon^4-12\tilde{\eta}+4\tilde{\eta}^2-4\epsilon^3(1+5\tilde{\eta})
-4\epsilon(3-3\tilde{\eta}+2\tilde{\eta}^2)+2\epsilon^2(1+6\tilde{\eta}+4\tilde{\eta}^2)}}{2(-1+\epsilon)^2}$.
Now, the small-argument limit of (\ref{eq:scalar2}) can be taken only if
\begin{equation}
\label{eq:scalar3}
\left|\frac{y}{2(1-\epsilon)}\right|<1.
\end{equation}
This coupled with the fact that $\epsilon<1$ for inflation - see \cite{beyondslowroll} - and that (\ref{eq:scalar3}) will still be satisfied at $y=1$ - the horizon crossing - tells us that $\epsilon<0.5$. One can in fact, retain the $\left(\frac{y}{-1+\epsilon}\right)^{-a}$ prefactor for continuing beyond $\epsilon=0.5$ up to $\epsilon=1$, by using the following identity that helps in the analytic continuation of $\ _0F_1(a;z)$ to regions $|z|>1$ (i.e. beyond (\ref{eq:scalar3}))- see \cite{Wolfram}:
\begin{eqnarray*}
\label{eq:identity}
& & \frac{\ _0F_1(a;z)}{\Gamma(a)}=-\frac{e^{\frac{i\pi}{2}\left(\frac{3}{2} - a\right)}z^{\frac{1-2a}{4}}}{\sqrt{\pi}}\Biggl[sinh\left(\frac{\pi i}{2}\left(\frac{3}{2}-a\right)-2\sqrt{z}\right)
\sum_{k=0}^{\left[\frac{1}{4}(2|b-1|-1)\right]}\frac{(2k+|a-1|-\frac{1}{2})!}
{2^{4k}(2k)!(|a-1|-2k-\frac{1}{2})!z^k}\nonumber\\
& & +\frac{1}{\sqrt{z}}cosh\left(\frac{\pi i}{2}\left(\frac{3}{2}-a\right)-2\sqrt{z}\right)
\sum_{k=0}^{\left[\frac{1}{4}(2|b-1|-1)\right]}\frac{(2k+|a-1|-\frac{1}{2})!}
{2^{4k}(2k+1)!(|a-1|-2k-\frac{1}{2})!z^k}\Biggr],
\end{eqnarray*}
if $a-\frac{1}{2}\in{\bf Z}$.}
The power spectrum of scalar perturbations is then given by:
\begin{equation}
\label{eq:scalar5}
P^{\frac{1}{2}}_R(k)\sim\left|\frac{u_k(y=1)}{z}\right|\sim\left|H^{(2)}_{\tilde{\tilde{\nu}}}
\left(\frac{1}{(\epsilon-1)}\right)\right|\frac{1}{\sqrt{\epsilon}},
\end{equation}
where $z\sim a\sqrt{\epsilon}$ (in $M_\pi=1$ units), and \[\tilde{\tilde{\nu}}\equiv \frac{\sqrt{9+9\epsilon^4-12\tilde{\eta}+4\tilde{\eta}^2-4\epsilon^3(1+5\tilde{\eta})
-4\epsilon(3-3\tilde{\eta}+2\tilde{\eta}^2)+2\epsilon^2(1+6\tilde{\eta}+4\tilde{\eta}^2)}}{2(-1+\epsilon)^2}.\]

The tensor perturbation modes $v_k(y)$ satisfy the following equation (See \cite{beyondslowroll}):
\begin{equation}
\label{eq:tensor1}
y^2(1-\epsilon)^2v^{\prime\prime}_k(y) + 2y\epsilon(\epsilon-\tilde{\eta})u_k^\prime(y)+ (y^2-(2-\epsilon))v_k(y)=0.
\end{equation}
Using arguments similar to ones given for scalar perturbation modes' solution, one can show that the solution to (\ref{eq:tensor1}) is given by:
\begin{equation}
\label{eq:tensor2}
v_k(y)\sim y^{\frac{1-\epsilon^2+2\epsilon(-1+\tilde{\eta})}{2(-1+\epsilon)^2}}H^{(2)}_{\frac{\sqrt{9+\epsilon^4+4\epsilon(-6+\tilde{\eta})-4\epsilon^3\tilde{\eta}
+2\epsilon^3(9-4\tilde{\eta}+2\tilde{\eta}^2)}}{2(-1+\epsilon)^2}}\left(\frac{y}{(\epsilon-1}\right).
\end{equation}
The power spectrum for tensor perturbations is given by:
\begin{equation}
\label{eq:tensor3}
P^{\frac{1}{2}}_g(k)\sim|v_k(y=1)|\sim\left|H^{(2)}_{\tilde{\nu}}\left(\frac{1}{-1+\epsilon}\right)\right|,
\end{equation}
where $\tilde{\nu}\equiv\frac{\sqrt{9+\epsilon^4+4\epsilon(-6+\tilde{\eta})-4\epsilon^3\tilde{\eta}
+2\epsilon^3(9-4\tilde{\eta}+2\tilde{\eta}^2)}}{2(-1+\epsilon)^2}$. Hence, the ratio of the power spectra of tensor to scalar perturbations will be given by:
\begin{equation}
\label{eq:r}
r\equiv\left(\frac{P^{\frac{1}{2}}_g(k)}{P^{\frac{1}{2}_R(k)}}\right)^2
\sim\epsilon\left|\frac{H^{(2)}_{\tilde{\nu}}\left(\frac{1}{(\epsilon-1)}\right)}{H^{(2)}_{\tilde{\tilde{\nu}}}
\left(\frac{1}{(\epsilon-1)}\right)}\right|^2,
\end{equation}which, for $\epsilon=0.0034, \tilde{\eta}\sim{10}^{-5}$ - a set of values which are realized with Calabi-Yau volume ${\cal V}\sim {10}^5$ and $D3$-instanton number $n^s=10$ for obtaining $f_{NL}\sim{10}^{-2}$ and are also consistent with ``freeze-out" of curvature perturbations at superhorizon scales (See (\ref{eq:freezeout})) - yields $r=0.003$.  One can therefore get a  ratio of tensor to scalar perturbations of ${\cal O}(10^{-2})$ in slow-roll inflationary scenarios in Swiss-Cheese compactifications. Further, one sees that the aforementioned choice of $\epsilon$ and $\tilde{\eta}$ implies choosing the holomorphic isometric involution as part of the Swiss-Cheese Calabi-Yau orientifolding, is such that the maximum degree of the genus-0 holomorphic curve to be such that $n^0_\beta\sim \frac{\cal V}{k^2 g_s^{\frac{5}{2}}}$, which can yield the number of e-foldings $N_e\sim {\cal O}(10)$ for $D3$-instanton number $n^s\sim {10}$ alongwith  the non-Gaussianties parameter $f_{NL}\sim{\cal O}(10^{-2})$ and tensor-to scalar ratio $r=0.003$.

The expression for the scalar Power Spectrum at the super horizon scales i.e. near $y=0$ with $a(y) H(y)=$ constant, is given as:
\begin{equation}
\label{eq:freezeout4}
P^{(\frac{1}{2})}_R(y)\sim\frac{(1-\epsilon)^{\tilde{\tilde{\nu}}}y^{\frac{3}{2}-\tilde{\tilde{\nu}}}}
{H^{\nu-\frac{3}{2}}a^{\nu-\frac{1}{2}}\sqrt{\epsilon}}\sim A H(y) y^{\frac{3}{2}-\tilde{\tilde{\nu}}}
\end{equation}
where $\nu={\frac{1-\epsilon^2+2\epsilon(-1+\tilde{\eta})}{2(-1+\epsilon)^2}}$ and $A$ is some scale invariant quantity. Using $\frac{d ln H(y)}{d ln y}\equiv\frac{\epsilon}{1-\epsilon} $, we can see that scalar power spectrum will be frozen at superhorizon scales, i.e., $\frac{d ln P^{\frac{1}{2}}(y)}{d ln y}=0$ if the allowed values of $\epsilon$ and $\tilde\eta$ parameters satisfy the following constraint:
\begin{equation}
\label{eq:freezeout}
\frac{d ln H(y)}{d ln y}+ \frac{3}{2}-{\tilde{\tilde{\nu}}}\equiv \frac{\epsilon}{1-\epsilon}+ \frac{3}{2}-{\tilde{\tilde{\nu}}}\sim 0
\end{equation}

The loss of scale invariance is parameterized in terms of the spectral index which is:
\begin{equation}
\label{eq:freezeout6}
n_R - 1\equiv \frac{d ln P(k)}{d ln k} = 3 - 2 {\rm Re}(\tilde{\tilde{\nu}})
\end{equation}
which gives the value of spectral index $n_R-1=0.014$ for the allowed values e.g. say $(\epsilon=0.0034,\tilde{\eta}=0.000034)$ obtained with curvature fluctuations frozen of the order $10^{-2}$ at super horizon scales. 

In a nutshell, for ${\cal V}\sim{10}^{5}$ and $n^s\sim 10$ we have $\epsilon\sim{0.0034}, |\eta|\sim {0.000034}, N_e\sim{17}, |f_{NL}|_{max}\sim {10^{-2}}, r\sim 4\times{10}^{-3}$ and $|n_{R}-1|\sim{0.014}$ with super-horizon freezout condition's violation of ${\cal O}(10^{-3})$. Further if we try to satisfy the freeze-out condition more accurately, say we take the deviation from zero of the RHS of (\ref{eq:freezeout}) to be of ${\cal O}(10^{-4})$ then the respective set of values are: ${\cal V}\sim{10}^{6}$, $n^s\sim10$, $\epsilon\sim{0.00028}, |\eta|\sim {10}^{-6}, N_e\sim{60}, |f_{NL}|_{max}\sim {0.01}, r\sim{0.0003}$ and $|n_{R}-1|\sim{0.001}$. This way, we have realized $N_e\sim60$, $f_{NL}\sim {10}^{-2}$, $r\sim {10}^{-3}$ and an almost scale-invariant spectrum in the slow-roll case of our LVS Swiss-Cheese Calabi-Yau orientifold setup.

\section{Conclusion and Discussion}

In this note, we argued that starting from large volume compactification of type IIB string theory involving orientifolds of a two-parameter Swiss-Cheese Calabi-Yau three-fold, for appropriate choice of the holomorphic isometric involution as part of the orientifolding and hence the associated Gopakumar-Vafa invariants  corresponding to the maximum degrees of the genus-zero rational curves , it is possible to obtain $f_{NL}$ - parameterizing non-Gaussianities in curvature perturbations - to be of ${\cal O}(10^{-2})$ in slow-roll and to be of ${\cal O}(1)$ in slow-roll violating scenarios. Using general considerations and some algebraic geometric assumptions as above, we show that requiring a ``freezeout" of curvature perturbations at super horizon scales, it is possible to get tensor-scalar ratio of ${\cal O}(10^{-3})$ in the same slow-roll Swiss-Cheese setup. We predict loss of scale invariance to be within the existing experimental bounds. In a nutshell, for Calabi-Yau volume ${\cal V}\sim{10}^{6}$ and $n^s\sim 10$, we have realized $\epsilon\sim{0.00028}, |\eta|\sim {10}^{-6}, N_e\sim{60}, |f_{NL}|_{max}\sim {0.01}, r\sim{0.0003}$ and $|n_{R}-1|\sim{0.001}$ with a super-horizon-freezout condition's deviation (from zero) of ${\cal O}(10^{-4})$. Further we can see that with Calabi-Yau volume ${\cal V}\sim{10}^{5}$ and $n^s\sim 10$ one can realize better values of non-Gaussienities parameter and ``r" ratio ($|f_{NL}|_{max}=0.03$ and $r=0.003$) but with number of e-foldings  less than $60$. Also in the slow-roll violating scenarios, we have realized $f_{NL}\sim {\cal O}(1)$ with number of e-foldings $N_e\sim 60$ without worrying about the tensor-to-scalar ratio and $|n_R-1|$ parameter.

To conclude, we would like to make some curious observations pertaining to the intriguing possibility of dark matter  being modelled by the NS-NS axions presenting the interesting scenario of unification of inflation and dark matter and producing finite values of non-Gaussianities and tensor-scalar ratio. In (\ref{eq:nonpert21}), if one assumes:

\noindent (a) the degrees $k_a$'s of $\beta\in H_2^-(CY_3)$ are such that they are very close and large which can be quantified as $\frac{k_1^2-k_2^2}{k_1^2+k_2^2}\sim-{\cal O}\left(\frac{1}{2\sqrt{ln \cal V}{\cal V}^4}\right)$,

\noindent (b) one is close to the locus $sin(nk.b+mk.c)=0$, where the closeness is quantified as $sin(nk.b+mk.c)\sim{\cal O}(\frac{1}{\cal V})$, and

\noindent (c) the axions are sub-Planckian so that one can disregard quadratic terms in axions relative to terms linear in the same,

\noindent then the potential of (\ref{eq:nonpert21}) can then be written as
\begin{equation}
\label{eq:DM1}
V\sim V_0\left(\left(\sum_{m^a}e^{-\frac{m^2}{2g_s} + \frac{m_ab^a n^s}{g_s}}\right)^2 - 8\right).
\end{equation}
Now, the Jacobi theta function (``$\theta(\frac{i}{g_s},\frac{n^sb^a}{g_s})$") squared in (\ref{eq:DM1}) can be rewritten as:
\begin{equation}
\label{eq:DM2}
\sum_{{\cal M}_1^+,{\cal M}_1+^-;{\cal M}_2^+,{\cal M}_2^-}e^{-\frac{({\cal M}_1^+)^2+({\cal M}_1^-)^2}{2g_s}}e^{-\frac{({\cal M}_2^+)^2+({\cal M}_2^-)^2}{2g_s}}
e^{({\cal M}_1^+b^1+{\cal M}_2^+b^2)\frac{n^s}{g_s}}.
\end{equation}
Now, writing $m_1b^1+m_2b^2$ as $\frac{1}{2}({\cal M}_+(b^1+b^2)+{\cal M}_-(b^1-b^2))$ and noting that the inflaton ${\cal I}$, for $k_1\sim k_2$ can be identified with $b^1+b^2$ - see \cite{axionicswisscheese} - one sees that (\ref{eq:DM2}) can be written as
\begin{equation}
\label{eq:intermed}
\left(\sum_{{\cal M}_1^-}e^{-\frac{({\cal M}_1^-)^2}{4g_s}}\right)^2
\sum_{{\cal M}_+,{\cal M}_-}e^{-\frac{({\cal M}_+)^2+({\cal M}_-)^2}{2g_s}}e^{\frac{({\cal M}_+{\cal I} + {\cal M}_-{\cal I}^\perp) n^s}{2}},
\end{equation}
 ${\cal I}^\perp\sim b^1-b^2$ - for orthonormal axionic fields, ${\cal I}^\perp$ will be orthogonal to ${\cal I}$.  Now, assuming ${\cal I}^\perp$ has been stabilized to 0,
one sees that one could write
\begin{equation}
\label{eq:DM3}
\left(\theta\left(\frac{i}{g_s},\frac{b^an^s}{g_s}\right)\right)^2\sim2\left(\sum_{{\cal M}_1^-}e^{-\frac{({\cal M}_1^-)^2}{2g_s}}\right)^3\sum_{{\cal M}_+\geq0}e^{-\frac{{\cal M}_+^2}{2g_s}}cosh\left(\frac{{\cal M}_+{\cal I}n^s}{2}\right),
\end{equation}
which in the weak coupling limit $g_s<<1$ is approximately equal to $2\sum_{{\cal M}_1\geq0}e^{-\frac{{\cal M}_1^2}{4g_s}}cosh\left(\frac{{\cal M}_+{\cal I}n^s}{2}\right)$. This, when substituted into the expression for the potential in (\ref{eq:DM1}), yields:
\begin{equation}
\label{eq:DM4}
V\sim V_0\left(\sum_{{\cal M}\geq0}e^{-\frac{{\cal M}^2}{2g_s}}cosh\left(\frac{{\cal M}{\cal I}n^s}{2}\right) - \sum_{{\cal M}\geq0}e^{-\frac{{\cal M}^2}{2g_s}}\right).
\end{equation}
Once again, in the weak coupling limit, the sum in (\ref{eq:DM4}) can be assumed to be restricted to
${\cal M}$ proportional to 0 and 1. This hence gives:
\begin{equation}
\label{eq:DM5}
V\sim V_0\left(cosh\left(\frac{{\cal I}n^s}{2}\right) - 1\right).
\end{equation}
One sees that (\ref{eq:DM5}) is of the same form as the potential proposed in \cite{SahniWang}:
\[V=V_0\left(cosh(\lambda\phi)-1\right),\]
for cold dark matter! This, given the assumption of sub-Planckian axions, is by no means valid for all ${\cal I}$. However, in the given domain of validity, the fact that a string (SUGRA) potential can be recast into the form (\ref{eq:DM5}) is, we feel, quite interesting.

Alternatively, in the same spirit as \cite{gravwavesKallosh}, if one breaks the NS-NS axionic shift symmetry ``slightly" by restricting the symmetry group ${\bf Z}$ to ${\bf Z}_+\cup\{0\}$, then (\ref{eq:DM4}) can be rewritten as:
\begin{equation}
\label{eq:Q1}
\sum_{{\cal M}\geq0}e^{-\frac{{\cal M}^2}{2g_s}}e^{\frac{{\cal M}{\cal I}n^s}{2}} - \sum_{{\cal M}\geq0}e^{-\frac{{\cal M}^2}{2g_s}}\sim e^{-\frac{\pi^2}{2g_s}}e^{\frac{\pi{\cal I}n^s}{2}}
+ e^{-\frac{4\pi^2}{2g_s}}e^{\frac{4\pi{\cal I}n^s}{2}},
\end{equation}
which is similar to:
\[V=e^{\alpha_1+\alpha_2\phi}+e^{\beta_1+\beta_2\phi},\]
(where $\alpha_2,\beta_2$ are taken to be positive) that has been used to study quintessence models (in studies of dark energy) - see \cite{Q} -  in fact, as argued in \cite{Q}, one can even include ${\cal I}^\perp$.

\section*{Acknowledgements}

The work of PS was supported by a CSIR Junior Research Fellowship. The work of AM was partly supported by the Perimeter Institute, Canada and the Abdus Salam International Centre for Theoretical Physics, Italy. AM also thanks the theory groups of Cornell University (especially Henry Tye), Caltech, UCLA and UC Berkeley (especially Ori Ganor) for the warm hospitality where part of this work was completed - AM also thanks Henry Tye for useful discussions and especially S.Sarangi and W.Kinney for very useful clarifications.

\end{document}